# Thermal rectification and interface thermal resistance in hybrid pillared-graphene and graphene: A molecular dynamics approach

Farrokh Yousefi,[a] Farhad Khoeini,[*a] Ali Rajabpour[b]

[a]Department of Physics, University of Zanjan, Zanjan, 45195-313, Iran

[b]Advanced Simulation and Computing Lab. (ASCL), Mechanical Engineering Department, Imam Khomeini International University, Qazvin, 34148–96818, Iran

Corresponding author: Farhad Khoeini

## Abstract

In this study, we investigate the thermal rectification and thermal resistance in the hybrid pillared-graphene and graphene (PGG) system. This is done through the classical molecular dynamics simulation (MD) and also with a continuum model. At first, the thermal conductivity of both pillared-graphene and graphene is calculated employing MD simulation and Fourier's low. Our results show that the thermal conductivity of the pillared-graphene is much smaller than the graphene by an order of magnitude. Next, by applying positive and negative temperature gradients along the longitudinal direction of PGG, the thermal rectification is examined. The MD results indicate that for the lengths in the range of 36 to 86 nm, the thermal rectification remains almost constant (~3-5%). We have also studied the phonon density of states (DOS) on both sides of the interface of PGG. The DOS curves show that there is phonon scattering at low frequencies (acoustic mode) that depends on the imposed temperature gradient direction in the system. Therefore, we can introduce the PGG as a promising thermal rectifier at room temperature. Furthermore, in the following of this work, we also explore the temperature distribution over the PGG by using the continuum model. The results that obtained from the continuum model predict the MD results such as the temperature distribution in the upper half-layer and lower full-layer graphene, the temperature gap and also the thermal resistance at the interface.

## 1. Introduction

Graphene[1,2], which is a two-dimensional honeycomb form sheet, has awesome properties such as high mechanical elastic modulus[3,4], superior thermal conductivity[5–8] and electrical properties[9–13]. Owing to these exceptional properties, cause that most people around the world try to employ graphene and graphene-based materials in various fields[14]. For example, to manage heat in the nano-size devices, the graphene can be a good candidate because of its excellent thermal conductivity. Another interesting subject is an investigation of the interface of graphene and other materials to gain various properties. In Ref [15], the interfacial thermal resistance between graphene and hydrogenated graphene (graphane) were examined. They found that in such a system there is a preferred direction for heat current. The heat current passes through the system from graphane to graphene. This is the so-called thermal rectifier in analogy to an electrical diode. In the thermal rectifiers, the magnitude of heat flux significantly depends on the sign of applied temperature gradient bias. This phenomenon had also been seen in other hybrid systems[16–21].

Thermal rectification is an interesting issue that is used in phononic systems. Zhang *et al.*[22] constructed a "Y" shape system from graphene, and by using molecular dynamics approach explored the thermal rectification. They demonstrated that heat flux in that system runs from branches to the stem, therefore the system was introduced as a promising thermal rectifier. Thermal rectification also has been seen in asymmetric graphene nanoribbon[23], carbon nanocone[24], and radial graphene[25]. Thermal rectification has a wide range of applications in heat management, information processing[26] or thermal logic circuits[27,28]. For example, Pal *et al.*[29] used monolayer graphene nanoribbon to build a thermal "AND" gate, which performs logic calculations with phonons, using two asymmetric graphene nanoribbons. These asymmetric graphene blocks play as thermal diodes. In the next step, the thermal diodes are employing to construct thermal transistors same as its electronic counterpart[30].

Moreover, In Ref.[31] a hybrid system contains a bilayer and a single layer was investigated through molecular dynamics simulation and continuum modeling to obtain thermal properties of the system. In the current paper, we try to study interface thermal resistance and also thermal rectification in hybrid pillared-graphene and graphene by using non-equilibrium molecular dynamics simulation (NEMD). The pillared-graphene is a three-dimensional carbon-based architecture that was constructed with carbon nanotubes and graphene sheets. Recently, the thermal conductivity of the pillared-graphene network was calculated using NEMD along the parallel and perpendicular directions concerning graphene sheet[32]. We indicate that thermal resistance at the interface between two parts of the system (pillared-graphene and graphene) is significantly large and depends on the imposed temperature gradient direction. The thermal resistance behavior with respect to the temperature gradient leads to prefer direction for heat flux. Therefore, we introduce a device as a promising thermal rectifier that can be used as a primitive block in the phononic instruments.

## 2. Computational method

In this work, all of non-equilibrium molecular dynamics simulations were performed using classical LAMMPS package[33]. The simulated system is hybrid pillared-graphene and graphene with a width of 66 Å and length of $L$ which can be varied in different simulations. This pillared-graphene is constructed by two graphene sheets in this system. The schematic view of the sample illustrated in Fig. 1. The distance between two graphene sheets (pillar length) in the pillared-graphene is ~16 Å and the chirality of carbon nanotubes is (6, 6)[32]. Also, the distance between two vertical nanotubes (inter-pillar distance) is ~36 Å. At the interface of the vertical carbon nanotube and graphene, a vacancy with a type of 6-7-6-7 is created. Periodic boundary condition was applied along the in-plane direction and free boundary condition was assumed for the out-plane direction. To describe the interaction between carbon atoms in this system, we used optimized three-body Tersoff potential[34]. The Tersoff potential which is appropriate for modeling covalent bonds allows breaking or creation of bonds. The time step was considered 1 fs to integrate the equations of motion by employing the velocity Verlet algorithm.

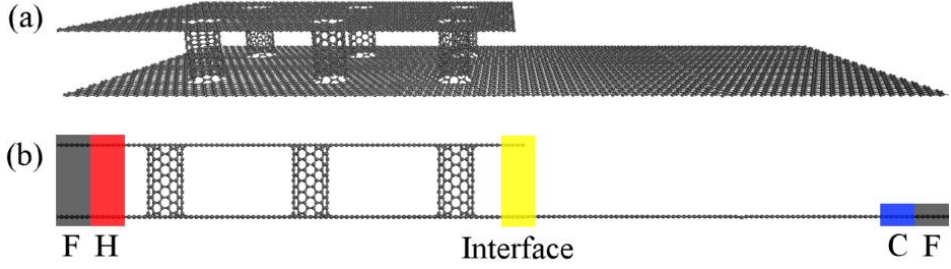

FIG. 1. (a) Perspective view and (b) side view of the PGG. The atoms in the black regions (shown with *F*) are fixed during simulation time. Red and blue-colored regions (labeled with *H* and *C*) also indicate the hot and cold baths, respectively.

For eliminating the extra stress of sample and relaxing the system, at first the PGG was implemented under the NPT ensemble via Nosé-Hoover[35] thermostat and barostat for 5 ns. In this stage, the stress along in-plane directions will reach zero and the temperature of the sample will be 300 K. After the relaxation process, the system as shown in Fig. 1 was imposed positive and a negative temperature gradients along the longitudinal direction of the sample for during 40 ns. Two temperature gradients bias were applied because of heat current magnitude significantly depends on the sign of temperature gradient bias. Thermal rectification quantity is obtained by using the equation below[8,25],

$$TR\% = \frac{j_n - j_p}{j_p} \times 100 \tag{1}$$

where $j_n$ and $j_p$ are the heat current in positive and negative directions, respectively. Also, we have examined the phonon density of states (DOS) on both sides of the interface to investigate the phonon scattering from the interface. The phonon scattering leads to thermal resistance at the interface. For calculating DOS, we carried out Fourier's transform on velocity autocorrelation function[36,37] (illustrated by $< >$),

$$P(\omega) = \frac{m}{k_B T} \int_0^\infty e^{-i\omega t} < \vec{v}(0).\vec{v}(t) > dt \qquad (2)$$

where $m, k_B, T$ are atomic mass, Boltzmann constant, and the temperature, respectively. Also, the parameters $\omega, \vec{v}, t$ are phonon angular frequency, velocity and, simulation time, respectively.

It is notable that the thermal conductivity of two parts of the system with a length of 33 nm (*L*/2), i.e. pillared-graphene and also graphene, were calculated in the same way as discussed above. The thermal conductivity of the pillared-graphene was obtained about 44 W/mK and for the grapheme was about 577 W/mK, which are in agreements with pillared-graphene of type PL32_CNT_MIPD21 in Ref.[32], and with graphene in Ref.[38].

Moreover, for calculation of the temperature distribution in the systems such as telescopic silicon nanowire[17] or bilayer/monolayer graphene heterostructure [31] in the long scale, there are some useful models upon heat conduction equation. After stating MD results, we employed a continuum model that describes temperature distribution in the long PGG system. The continuum model adapted from Ref.[31] to predicted the local temperature in our system using the heat equation.

## 3. Results and discussion

All of NEMD simulations were implemented to get the results of thermal rectification in hybrid PGG at room temperature. All simulations were performed for 40 ns. We selected the last 20 ns and averaged the temperature values to plot the temperature profile which is shown in Fig. 2 (for two cases of positive and negative temperature gradient directions). The temperature drop (gap) is observed across the interface of PGG. The temperature gap shows that there is a significant scattering of phonons passing through the interface. The temperature gap indicates that there is a thermal resistance between pillared-graphene and graphene. The thermal resistance (R) is related to the heat flux with the equation below,

$$R = \frac{\Delta T_{gap}}{j} \qquad (3)$$

The thermal resistance in analogy to electrical resistance prevents more phonon transmission across the interface.

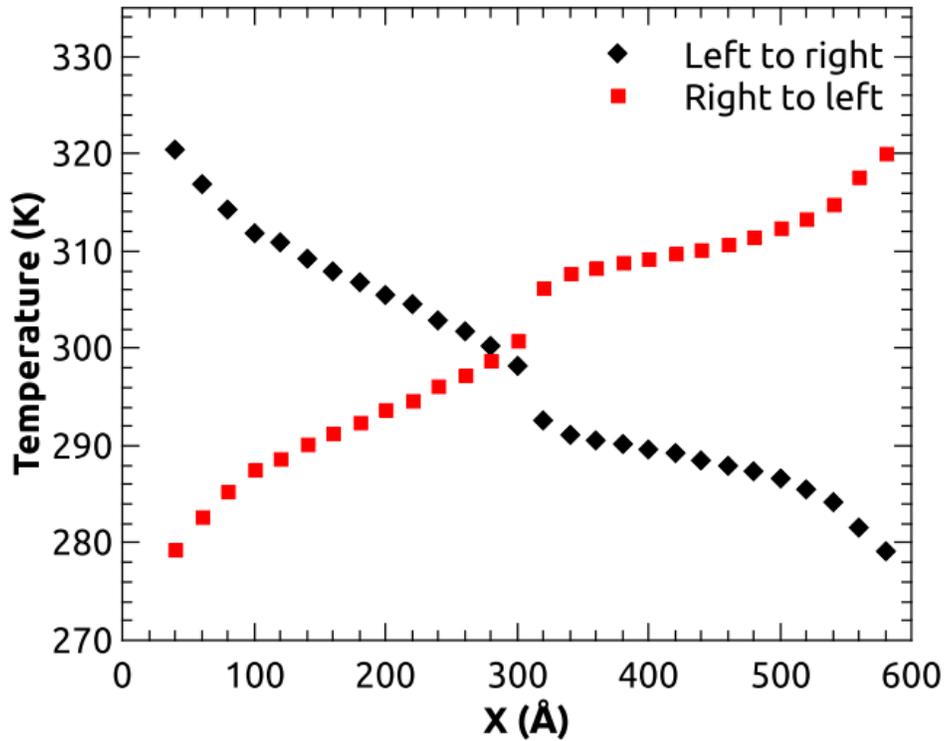

FIG. 2. The temperature profile in the PGG for two opposite temperature gradient directions. Black and red-colored curves indicate left to right and right to left temperature gradient directions, respectively.

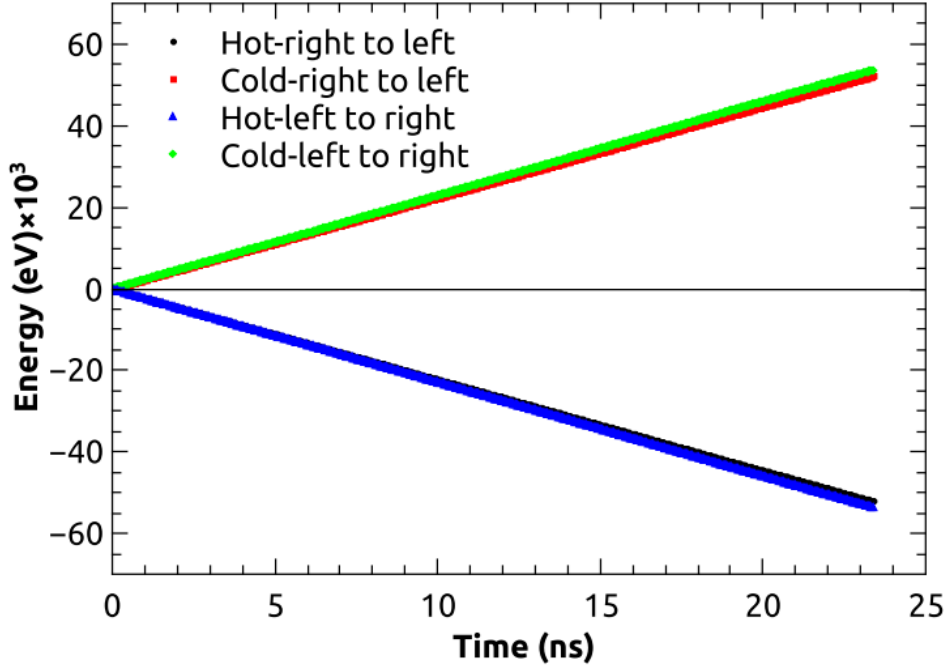

FIG. 3. The accumulative energy extracted and removed from the hot and the cold baths, respectively. The slope of the energy is the heat current that flows along the temperature gradient. Green and blue-colored curves are for the right-to-left gradient while that red and black-colored are for the opposite direction.

In order to explore thermal rectification in the PGG, it needs to calculate the heat current that flows in the sample along with two opposite directions by reversing the temperature gradient. According to Fig. 3, we obtained the accumulative energy extracted from the hot bath or removed from the cold one. The slope of the curves indicates the amount of heat current that flows in the system. As shown, heat current that flows from the left to the right side is greater than from the opposite direction. This means that the heat current prefers one direction to another direction. This phenomenon occurs when the mass distribution of the system is non-symmetric[8,16,39–41]. Here, half of the system is heavier than the other half part. In the non-symmetric systems, it is usually preferable that the heat current flows from the heavier part to the lighter part. Therefore, in our setup, the heat current transfers from the pillared-graphene to the graphene more than from the opposite direction. The fundamental reason for thermal rectification is related to phonon scattering from the interface that we will discuss below. As a result, the calculated heat currents in both directions are plotted in Fig. 4 for different lengths of the system. In all lengths, the heat current that flows from the left to right direction is greater than from the opposite direction which is the concept of thermal rectification. As we expect, the heat current decreases with increasing sample length for a constant temperature difference (40 K) between the hot and the cold baths. The decreasing behavior of the heat current is slower than a linear behavior which is due to the phonon ballistic transport in small sample length[42].

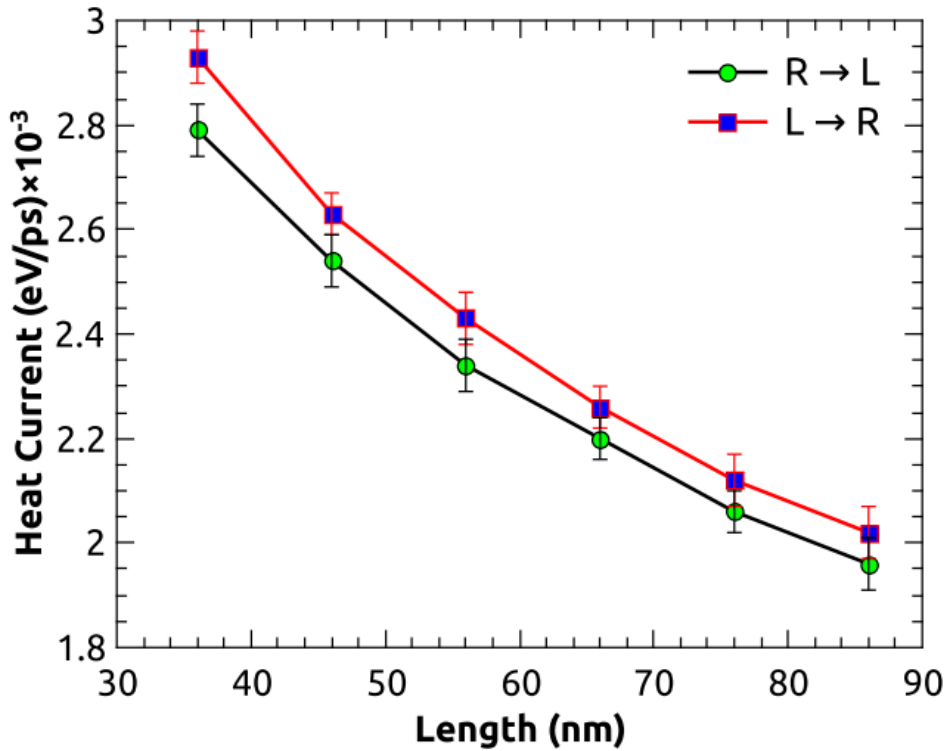

FIG. 4. The calculated heat current in both directions i.e. from the left side to the right one and vice versa. The heat current that flows from left to right is more than the opposite direction due to non-symmetric geometry and mass distribution.

To examine the thermal rectification, we have considered Eq. 1. As illustrated in Fig. 5, the thermal rectification is ~ 2–6% for sample length between 36 and 86 nm. For comparison, the obtained thermal rectifications in the previous works are ~20% [15], 0-15% [17], < 40% [41], 15-65% [25]. As already reported, the thermal rectification is not almost sensitive to the sample length but is sensitive to the temperature difference between the hot and the cold regions.

To investigate fundamentally the underlying mechanism of thermal rectification, we look at the phonon scattering at the interface. When phonons enter to the interface between pillared-graphene and graphene, some of them will be scattered due to non-symmetric of mass distribution. Therefore, a thermal resistance will appear which causes to decrease the heat current. On the other hand, the number of phonons that scatter from the interface strongly depends on the transport direction of the phonons. This means that the heat current experiences more resistance when the temperature gradient is from the narrow part to the thick than the case in which the temperature gradient is in the opposite direction. To show this issue, the phonon density of states (DOS) was calculated (according to Eq. 2) on both sides of the interface for the positive and negative temperature gradients. As indicated in Fig. 6, the DOSs on both sides of the interface is not completely overlapped. This states that some of the phonons could not pass through the interface to the other part due to interface thermal resistance. For both of the positive and negative temperature gradients, we see different interface thermal resistance from DOSs which are in agreement with the above discussion.

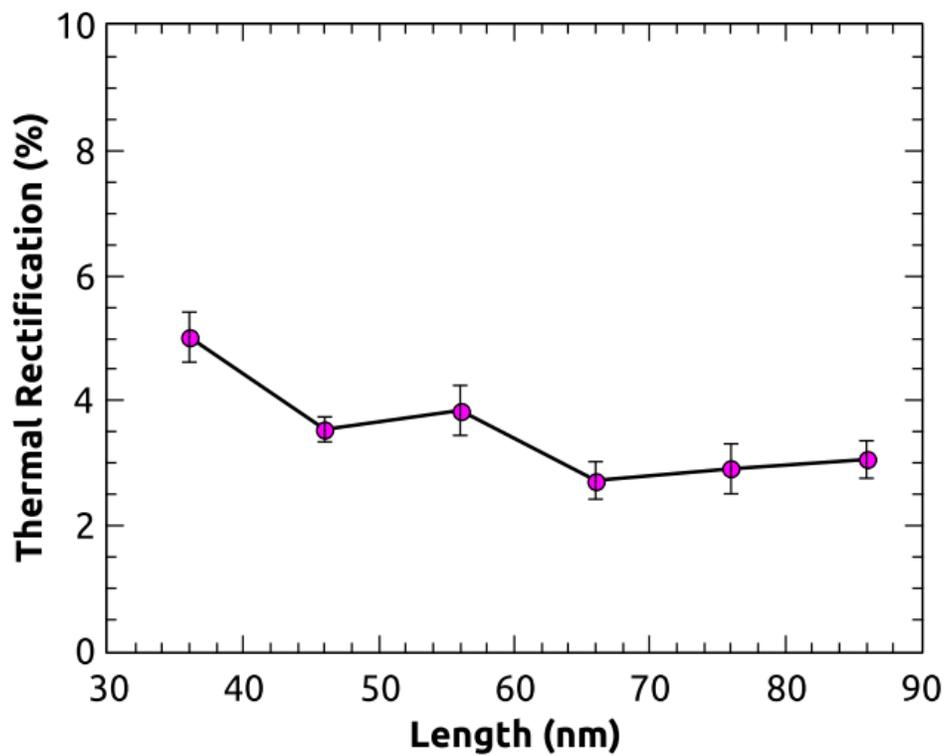

FIG. 5. Thermal rectification in the hybrid pillared-graphene and graphene. The range of sample length is 36 to 86 nm. Phonons are found to preferentially flow from the thick to the narrow part.

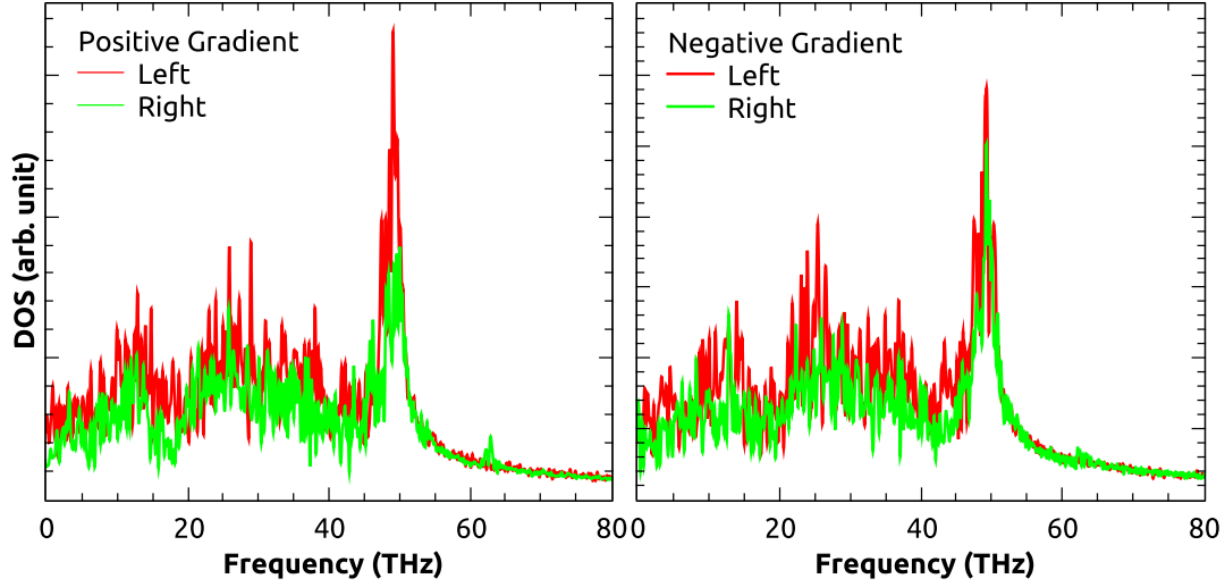

FIG. 6. The DOSs were calculated on both sides of the PGG interface for two cases of positive gradient and negative gradients. Those portions of the DOS that has no overlap with another curve (around 50 THz in positive gradient and ~25 THz for negative one) indicate phonon scattering in the interface.

## Continuum model

We intend to state here a continuum model according to Ref.[31] that can describe temperature distribution in a long PGG (i.e. $\gg$ 36Å). In this condition, we can suppose the pillared-graphene as two parallel graphene sheets (without defects) that was connected with a few numbers of carbon nanotubes. By an approximation, we can assume this system is the same as the bilayer/monolayer graphene heterostructure. The in-plane thermal conductivity of each graphene sheet was considered as $\kappa_{ip}$ and the cross-plane thermal conductivity was considered as $\kappa_{cp}$. The $\kappa_{cp}$ is equal to the number of single-walled carbon nanotubes (SWCN) in pillared-graphene multiplied by their thermal conductivity ($\kappa_{cp} = n \times \kappa_{SWCN}$).

Considering the above assumptions, we can solve the heat conduction equation along the PGG. The temperature distribution in the upper half-layer graphene (*hlg*) and also lower full-layer graphene (*flg*) were extracted as follows:

$$\frac{d^2 T_{hlg}(x)}{dx^2} - P\big(T_{hlg}(x) - T_{flg}(x)\big) = 0, \qquad 0 \leq x \leq L/2 \qquad (4.a)$$

$$\frac{d^2 T_{flg}(x)}{dx^2} - P\big(T_{hlg}(x) - T_{flg}(x)\big) = 0, \qquad 0 \leq x \leq L/2 \qquad (4.b)$$

$$\frac{d^2 T_{flg}(x)}{dx^2} = 0, \qquad\qquad\qquad L/2 \leq x \leq L \qquad (4.c)$$

where $P = \frac{\kappa_{ip}}{\kappa_{cp} \times l^2}$ and $l$ is the pillar length. The boundary conditions for solving the above equations are:

$$x = 0 \rightarrow T_{flg} = T_{Hot}, \qquad (5.a)$$

$$x = L/2 \rightarrow dT_{hlg}/dx = 0, dT_{flg}/dx \text{ and } T_{flg} \text{ are continues functions.} \qquad (5.b)$$

$$x = L \rightarrow T_{flg} = T_{Cold}, \qquad (5.c)$$

where the $T_{Hot}$ and $T_{Cold}$ are the hot and cold temperatures, respectively. The two above mentioned equations are coupled and therefore we should solve them simultaneously. By doing a little math and algebraic operation, we find $T_{hlg}(x)$ and $T_{flg}$,

$$T_{hlg}(x) = T_{Hot} - B\left[\frac{x}{2} - \frac{\sinh(\sqrt{2P}x)}{2\sqrt{2P}\cosh(\sqrt{P/2}L)}\right], \qquad 0 \leq x \leq L/2, \qquad (6.a)$$

$$T_{flg}(x) = T_{Hot} - B\left[\frac{x}{2} + \frac{\sinh(\sqrt{2P}x)}{2\sqrt{2P}\cosh(\sqrt{P/2}L)}\right], \qquad 0 \leq x \leq L/2, \qquad (6.b)$$

$$T_{flg}(x) = T_{Hot} - \frac{T_{Hot} - T_{Cold}}{L}x, \qquad L/2 \leq x \leq L, \qquad (6.c)$$

where $B = \left(\frac{T_{Hot} - T_{Cold}}{L/2}\right) / \left(\frac{3}{2} + \frac{\tanh(\sqrt{P/2}L)}{L\sqrt{2P}}\right)$. By using Eqs. 6, we can obtain the average temperature of pillared-graphene (left side) near to the interface i.e. $\frac{\left(T_{hlg}\left(\frac{L}{2}\right) + T_{flg}\left(\frac{L}{2}\right)\right)}{2} = T_{Hot} - BL/4$ and also the temperature of the graphene (right side) i.e. $T_{flg}\left(\frac{L}{2}\right) = \frac{T_{Hot} + T_{Cold}}{2}$.

The difference between two temperatures on both sides of the interface shows a temperature drop at the interface which we can also see by NEMD simulation (Fig. 2) for a finite length of PGG:

$$\Delta T_{interface} = \frac{T_{Hot} - T_{Cold}}{2} - BL/4. \tag{7}$$

The purpose of exploring Eq. 7 is to show that there is a temperature gap and thermal resistance at the interface by employing the continuum model in a special situation (large inter-pillar distance) in a long PGG.

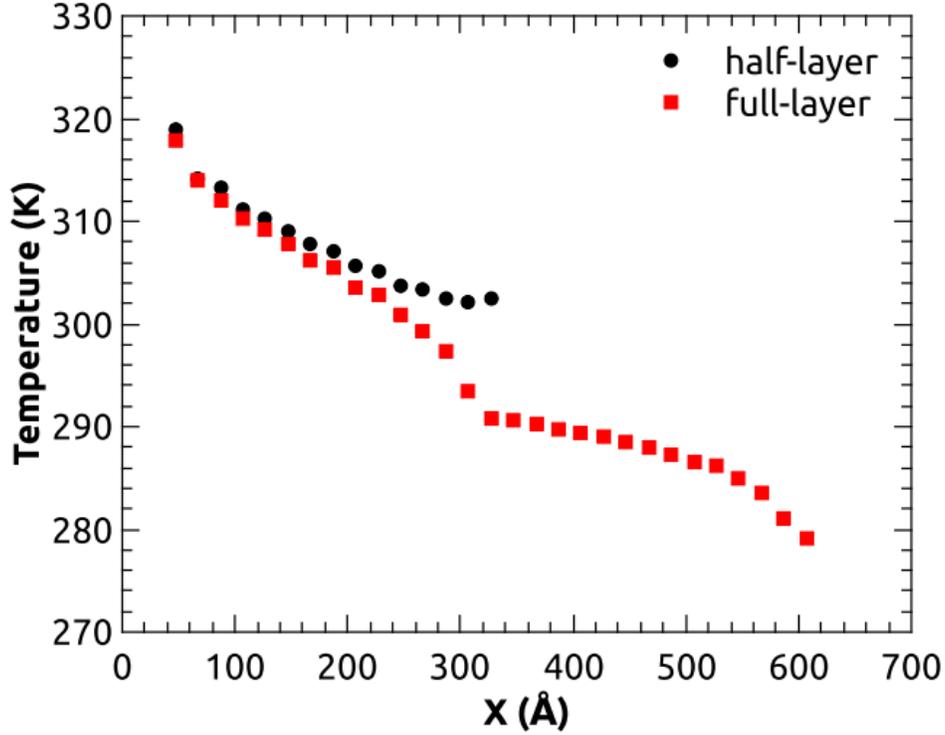

Fig. 7. The temperature profile of the upper half-layer graphene (black) and lower full-layer graphene (red) obtained from MD simulation.

Also, the Eqs. 6.a and 6.b show that there is a difference between temperatures of upper half-layer graphene and lower full-layer graphene at the interface for a long system and long inter-pillar distance,

$$T_{hlg}(L/2) - T_{flg}(L/2) = \frac{\sinh(\sqrt{2P}L/2)}{\sqrt{2P}\cosh(\sqrt{P/2}L)}. \tag{8}$$

According to Fig. 7, we also see that there is a temperature difference at the interface between *hlg* and *flg* by using MD simulation that confirms Eq. 8.

Therefore, we can state that the continuum model describes the temperature distribution across the system and predict the temperature gap at the interface of PGG. When we have a temperature gap in the temperature profile, a thermal resistance will appear according to Eq. 3.

## 4. Conclusions

Using a series of NEMD simulations, the thermal rectification of hybrid pillared-graphene and graphene were investigated. The influence of the sample length on the thermal rectification was examined. The thermal rectification values were in the range from 3 to 5% for the sample length of 36-86 nm. Our results show that the interface thermal resistance was responsible for the thermal rectification and phonons scattering. Also, we have studied the underlying mechanism of the thermal resistance and also thermal rectification by calculating the phonon density of states on both sides and near the interface. This quantity showed that there are not completely overlaps between two DOSs and therefore this issue leads to interfacial thermal resistance that significantly depends on the imposed temperature gradient direction. We have also indicated that the heat current decreases with an increasing sample length of PGG slower than a linear behavior which is in agreement with the ballistic transport concept. Finally, a continuum model has been presented to describe temperature distribution according to the heat conduction equation. This model predicts a temperature gap and a thermal resistance at the interface.

## Conflicts of interest

There are no conflicts to declare.

Email: khoeini@znu.ac.ir